\documentstyle[12pt,aasms4,epsf]{article}

\received{}
\accepted{ }

\begin{document}
\title {Magnetic field effects on the  
thermonuclear combustion front of Chandrasekhar
mass white dwarfs}

\author{Cristian R. Ghezzi\altaffilmark{1}, Elisabete M. de Gouveia Dal 
Pino\altaffilmark{1}
\& Jorge E. Horvath\altaffilmark{1} }
\altaffiltext{1}{Instituto Astron\^omico e Geof\'{\i}sico, University
of S\~ao Paulo, Av. Miguel St\'efano, 4200, S\~ao Paulo
04301-904, SP, Brasil;
E-mail: cghezzi@jet.iagusp.usp.br,  }

\begin{abstract}
The explosion of a type Ia supernova starts in a white dwarf 
as a laminar deflagration at the center of the star and soon several 
hydrodynamic instabilities, in particular, the Rayleigh-Taylor instability,
begin to act. A cellular stationary combustion 
and a turbulent combustion regime are rapidly achieved by the 
flame and maintained up to the end of the so-called flamelet regime when 
the transition to detonation is believed to occur. The burning velocity at these regimes is well 
described by the fractal model of combustion.
Using a semi-analytic approach, we describe the effect of magnetic fields 
on the fractalization of the front considering a white dwarf  
with a nearly dipolar magnetic field. We find 
an intrinsic asymmetry on the velocity field that may be maintained up
 to the free expansion phase of the remnant. 
Considering the strongest values inferred for a white dwarf's 
magnetic fields with  strengths up to $10^{8}-10^{9}$ G at the 
surface and assuming that the field near the centre is roughly 10 times greater,
asymmetries in the velocity field higher than $10-20 \%$ are
produced between the magnetic polar and the equatorial axis of the remnant which  
may be related to the asymmetries found from 
recent spectropolarimetric observations of very young
SN Ia remnants.
Dependence of the asymmetry with white dwarf composition is also 
analyzed.

\end{abstract}


\section{Introduction}

The explosion of a type Ia supernova begins with the combustion, 
at the center of a Chandrasekhar mass white dwarf of carbon-oxygen (C+O)  or 
oxygen-neon-magnesium (O+Ne+Mg) fuels. The heat is transported mainly 
by conduction due to degenerate and completely relativistic electrons 
as a subsonic deflagration wave propagatating outwardly of the star. 

The deflagration front born laminar is subject to several hydrodynamic instabilities 
such as Landau-Darrieus (LD) and Rayleigh Taylor (RT) instability
(Arnett \& Livne 1994, Khokhlov 1993) that produce an 
increment of the area at which the nuclear reactions take place.
This causes an increase of the nuclear energy generation rate and consequently an acceleration of the front. 
There are two scale ranges that can be distinguished,
at the smallest or ``microscopic" perturbation scales the LD predominates over 
the RT instability, while for 
the greatest scales ($\gtrsim 10^{5} \, {\rm cm} $) the RT instability is more 
important\footnote{This depends on the value of the Froude number $F=v_{lam}^{2}/gL$ 
(where $L$ and $v_{lam}$ are defined below), then if $F<1$ RT 
instability predominates over LD.}. 
The combustion front is stabilized by the merging of cells, the formation of cusps, and  
the expansion of the exploding star, which leads to the formation of a cellular 
structure at microscopic scales.  LD instabilities
lead to an aceleration no higher than  
$ 2 \% $ near the center of the star because they are nonlinearly stabilized (Khokhlov 1995). 
On the other hand, at the macroscopic scales there is a critical wavelength 
above which the nonlinear stabilization fails\footnote{At scales greater than this wavelength the Froude number is $F \ll 1$, and the fluid is fully turbulent.}  Zel'dovich et al. (1985).

The bubbles that grow due to RT instability are also subject to Kelvin-Helmoltz (KH) or 
shear instability when nonlinear stabilization fails. The onset of the KH instability marks the 
transition to the fully developed turbulence regime.
During this, fluid motions are characterized by the formation of a 
turbulent cascade in the inertial scales where viscous 
dissipation is not important. This turbulence can be described 
by the Kolmogorov's scaling law.

The acceleration of the turbulent thermonuclear flame due to the
action of the several hydrodynamics instabilities can be 
described by the fractal model first introduced  
by Woosley (1990), Timmes \& Woosley (1992), Niemeyer \& Hillebrandt (1995) and
 Niemeyer \& Woosley (1997). The  idea of applying fractals to the combustion fronts is useful because the statistic properties of the surface change self-similarly 
on the spatial scales. 
The wrinkled surface $\bar S$ behaves like a fractal 
$\bar S \propto \bar{R}^{D}$ , 
where $D$ is the fractal dimension of the surface, with
$2\leq D<3$, and $\bar{R}$ 
is the mean radius of the wrinkled surface (Filyand, 
Sivashinsky \& Frankel 1994). There is a scale range at which
the similarity holds, called the similarity range, that goes
from $l_{min}$ to $L$.
The increase of the area at which the nuclear reactions occur 
induces an increase in the propagation velocity that 
also behaves like a fractal\footnote{We refer here to the fractal 
dimension (and not to the Hausdorff-Besicovitch dimension) of an attractor as 
it is commonly defined in astrophysical works (Barnsley 1988). The 
wrinkled surface is not a fractal in the mathematical sense, it is 
converging to its attractor asymptotically with time, but it is 
possible to treat the surface as a fractal as long as 
$l_{min}/L \ll 1$, for ex.,
$l_{min}/L \simeq 10^{-5}$ at $\rho \simeq 10^{8}$ 
${\rm g\,cm^{-3} }$ (see below).} (see Niemeyer \& Hillebrandt, 1995).
This fractal model is supported by laboratory experiments
involving different gas mixtures (Gostintsev, Istratov \& Shulenin 1988). 

The propagation of the front subject to LD instability is well 
described by the Sivashinsky equation (Sivashinsky 1977, 1983) 
that also includes thermodiffusive and acceleration effects,
or by its gereralization (Frankel 1990).
Numerical studies of wrinkled surfaces utilizing 
these equations are able to reproduce the  
front on a wider range of spatial scales than direct hydrodynamic 
simulation and allow the study of the fractal properties of the 
flame (Blinnikov \& Sasorov 1996).
 These simulations of the Sivashinsky equation show that the
fractal growth of the front increases its velocity   
(Filyand, Sivashinsky \& Frankel 1994).

Several physical effects are yet to be considered for
a complete solution of the SN Ia problem. In particular, the effects of
magnetic fields which are known to be present in SN Ia progenitors
(believed to be white dwarfs in binary systems) have not been
considered. We will incorporate here the effects of magnetic fields on the 
fractal growth of the turbulent combustion front and show that they can 
break the spherical symmetry of the explosion.

\section{Formulation of the problem}
Ginsburg (1964) and Woltjer (1964) proposed that the magnetic
flux of a star is conserved during its evolution and collapse,
so that strong magnetic fields would be generated in degenerate
stars. Hence, a main sequence star with $R \sim 10^{10}$ cm 
and $B \sim 10-10^{5}$ G would collapse to form a 
white dwarf with $R \sim 10^{8}$ cm and 
$B \sim 10^{5}-10^{9}$ G on its surface.
Inferred magnetic field strengths of known magnetic
white dwarfs range from $\sim 10^{6}$  to  $ 10^{9}$ G (Jordan 1992). 
There is also an important percentage of known white dwarfs
in cataclysmic variables with magnetic fields ranging from $\sim 10^{6}$ to $10^{8}$ G
(see Chanmugan 1992, Liebert and Stockman 1983). There are several
models of white dwarfs with strong magnetic fields (Jordan 1992, Putney \& Jordan 1995,
Martin \& Wickramasinghe 1984), which assume centered dipolar, off-centered dipolar, 
quadrupolar or dipolar+quadrupole magnetic field configurations. Although the origin of the fields in these
 stars remains unclear it is frequently assumed that 
the fields are primordial.  In fact, 
Wendell, Van Horn \& Sargent (1987) carried out detailed calculation of 
the time evolution of a white dwarf's magnetic field and found the decay time of 
the field  is always longer than the typical cooling times 
of white dwarfs ($\tau \sim 10^{10}$ yr).
We will assume in this letter that the progenitor star of an SN Ia is a magnetic white 
dwarf with a centered dipolar magnetic field.  
Considering magnetic fields with the strongest inferred values, i.e., in the range $10^{8}-10^{9}$ G,
and using the model of Wendell, Van Horn \& Sargent (1987), we find 
 magnetic field strengths near the center of the star 
which are in the range $\sim 10^{9}-10^{10}$ G.
We thus must expect that when the explosion begins the flame 
will propagate parallel to the magnetic fields lines 
at the magnetic poles of the star, and perpendicularly 
to the field lines at the equator.

An estimate of the effective turbulent speed $u_{t}$ of the flame was obtained by Damk\"{o}hler (1939), who
proposed that $u_{t} \sim v(L)$, where $v(L)$ is the average of the
turbulent velocity fluctuations on the largest hydrodynamical scale $L$.
Later Karlovitz, Denniston, \& Wells (1951) derived a statistical approach based on turbulent velocity correlations that fits better with laboratory experiments. The formula simplifies to 
$u_{t}=\lbrack 2 u_{lam} v(L) \rbrack^{1/2}$ in the limit $v(L) \gg v_{lam}$, where $v_{lam}$ is the laminar velocity. 
Niemeyer \& Hillebrandt (1995) generalized those previous models and found
\begin{equation}
u_{t} (l)=v_{lam} \lbrack v(l) / v_{lam}  \rbrack ^{n}  
\end{equation}
where the exponent $n$ is arbitrary, with  $n=1$ for Damk\"ohler's model and  $n=1/2$ for the model of Karlovitz et al.. 
Although numerical simulations of thermonuclear flames are in good agreement with the Karlovitz et al. model 
(see Khokhlov 1993), Niemeyer \& 
Hillebrandt use their own version to derive the scaling law for the turbulent flame speed. 
 We will give here a slightly different derivation of the scaling law for $u_{t}$. 
The turbulent motions of the fluid can be described by the Kolmogorov's scaling law
for incompressible turbulent velocity fluctuations  which gives (Landau \& Lifshitz 1959)
\begin{equation}
\label{Kolm}
v(l)=v(L) \biggl( \frac{l}{L} \biggr) ^{1/3}
\end{equation} 
this scaling is valid in the inertial range ($\eta\ll l \ll L$) between the dissipation scale $\eta$ and the
length scale  $L$ at which the turbulent velocities are in equilibrium with the RT instability. 
So if we assume that, at the largest hydrodynamic scales which are
subject to the RT instability,
 the turbulent velocity is in equilibrium with the velocity of RT bubbles $v_{RT}(L)\sim v(L)$, 
where $v_{RT}(L) \propto (g L)^{1/2}$, we can use eq. (2) to obtain the velocity of the turbulent 
fluctuations at the lower scales
$v(l)=v_{RT}(l)(L/l)^{1/6}$. From this it is clear that $v(l)\geq v_{RT}(l)$ for $l \leq L$. 
Therefore, the turbulent cascade will dominate on all scales below the largest RT unstable wavelength
 (see also Niemeyer \& Hillebrandt 1995). Using the last formula and eq.
 (1), we obtain $u_{t}(L)=u_{lam}\lbrack v_{RT}(l)/u_{lam} \rbrack^{n} (L/l)^{n/6} $.
Following Timmes \& Woosley (1992),
 we will use as a lower cutoff the minimum length scale $l_{min}$ that can deform the laminar flame front.
 This length is given by the condition $v_{RT}(l_{min})=v_{lam}$. Therefore the scaling law for the turbulent flame speed is
\begin{equation}
u_{t}(L)=v_{lam} \biggl( \frac{L}{l_{min}} \biggr)^{n/2}
\end{equation}

In a fractal description of the turbulent flame propagation, eq. (3) gives the fractal velocity of the front
$v_{frac} \, = v_{lam} (L/l_{min})^{D-2}$ (Niemeyer \& Woosley 1997, 
Niemeyer \& Hillbrandt 1995). Comparing this with eq. (3), we see that the fractal excess is $D-2=0.5$ for 
$n =1$, and  $0.25$ for $n=1/2$, which is in agreement with the fractal 
dimensions inferred from  numerical studies of SN Ia  explosions (Blinnikov, Sasorov \& Woosley 1995).

In the equation above we must determine  $l_{min}$ and $L$.
As we have seen, the characteristic RT velocity of the
growing modes must be larger than or equal to the laminar deflagration
velocity $v_{lam}$, or 
$v_{RT} (l_{min}) = v_{lam}$.
This implies that
$l_{min} \,n_{RT}(l_{min})= v_{lam}$, where 
$n_{RT}(l)$ is the inverse of the characteristic RT time
$n_{RT}= (1/ 2\pi)  \sqrt{g k \Delta \rho/ 2\rho}$. This results
\begin{equation}
\label{lambdapol}
l_{min}= 8\pi \biggl( \frac{ \frac{1}{2} 
\rho\, v_{lam}^{2}}{g \,\Delta \rho} \biggr) =l_{pol}
\end{equation}  
where $g$ is the 
acceleration of the gravity, $\Delta \rho = \rho_{u}-\rho_{b}$ is the difference 
between the densities $\rho_{u}$ 
and $\rho_{b}$ of the unburned and burned fuel, respectively, and 
$\rho=(\rho_{u} + \rho_{b})/2$. We denote $l$
as $l_{pol}$ hereafter, where the subscript $``pol"$
stands for polar since it gives 
the value of $l_{min}$ at the magnetic poles of the
star where the propagation of the flame is parallel 
to the magnetic field lines, and therefore unaffected by them.

The value of $L$ has been determined from the minimum  
between the maximum spatial extent of the density inversion 
and the radius of the flame $R_{f}$. The density 
inversion is due to electron
capture in an isobaric nuclear statistical equilibrium environment
that causes the number of electrons to decrease and the 
density to increase. We here use 
$L=R_{f}$ because, for densities 
$\lesssim 10^{9}$ ${\rm g\,cm^{-3}}$, the density inversion scale is 
greater than the stellar radius (Timmes \& Woosley 1992). The 
highest unstable mode $L$ has
the highest growing time since
$t_{RT}(l)=(4 \pi l/g)^{1/2}\,(\rho/\Delta \rho)^{1/2} \leq t_{RT}(L)$, for
$l_{pol} \leq l \leq L$. For
example, near the center of the star 
$L \sim 3 \times 10^{7}\,{\rm cm}$ and  
the perturbation takes $t_{RT}(L)\sim0.5\,{\rm s}$ to 
grow to an amplitude of the order of its wavelength.
 On the other hand, for $l_{pol} \simeq 10^{5}\,{\rm cm}$  
$t_{RT}(l_{pol})=0.03\,{\rm s}$ and the perturbation 
grows faster. Thus with a deflagration velocity $v_{lam} \sim 10^{7}\,{\rm cm\,s^{-1}}$, the flame must travel only $\sim 5\times 10^{6} \, {\rm cm}$ for the slowest perturbations, and $\sim 3 \times 10^{5}\,{\rm cm}$ for the fastest ones 
in order to the perturbations to grow to amplitudes of the order of their lengths. 
This suggest that all the perturbation scales grow very fast in the combustion front
and therefore, they are all important.

Our analysis is applicable only for densities $\rho \geq 10^{7}\,{\rm g\,cm^{-3}}$, 
since the turbulent motions will destroy the corrugated flamelet regime
for lower densities (Niemeyer \& Woosley 1997, Hillebrandt \& Reinecke 2000).

\section{Asymmetric velocity field}
Near the center of the star the magnetic pressure ($B^{2}/8\pi \sim 10^{18}\,{\rm erg\,cm^{-3}}$) is much smaller than the gas pressure behind the front ($p \sim 10^{27}\,{\rm erg\,cm^{-3}}$). 
Therefore, it is reasonable to assume that the scaling law for the 
flame speed (eq. 3) will not be modified by the presence of the magnetic field.
This means that the fractal dimension can be considered independent of the 
magnetic field. 
On the other hand, the presence of magnetic fields will shift the lower unstable scales at the equator of the star with respect to the poles, as we will see, 
thus producing different fractal velocities at the magnetic poles and the equator 
of the star. 
We note also that
the turbulent fluid motions can only scramble the field inside the
``flame brush", that is, inside the RT foam that mixes unburned and burned material, 
or behind the front, and therefore,  they will not affect the large scale magnetic field geometry of the progenitor.

The dispersion relation for the RT instability
of the front with a magnetic field transverse to the direction
of the gravity and to the propagation of the flame, in a heterogeneous inviscid
plasma of zero resistivity is (Chandrasekhar 1961):
\begin{equation}
n_{RTB}= \frac{1}{2 \pi} \sqrt{g\,k\,\biggl( \frac{\Delta \rho}{2 \rho}-
\frac{k\,B^{2}}{4 \pi g \rho} \biggr) }
\end{equation}
here  $k= 2 \pi / l$, and $B$ is the magnetic field strength.
It is, therefore, possible to define an effective surface tension,
$T_{eff}=B /(2 \pi k) $, for which the magnetic field will have a stabilizing effect 
over perturbations with wavelengths below $\sim B^{2}/(g \Delta \rho)$. 
However, if the velocity of the flame is taken into account another
cutt-off length appears, $l_{eq}$, which is given by the condition
$l_{eq} n_{RTB}(l_{eq})=v_{lam}$. From this equation the
minimum scale of the self-similar range in presence of magnetic field is
\begin{equation}
\label{lambdaeq}
l_{eq}= \frac{8 \pi (\frac{B^2}{8 \pi}+
\frac{1}{2} \rho \, v_{lam}^{2} )}{g \Delta \rho}
\end{equation}  
here $eq$ stands for ``equator" because $l_{eq}$ gives  
the minimum instability scale at the magnetic equator of the star where 
the propagation of the front is perpendicular to the field lines.
Inserting eq. (6) in the definition of
the fractal velocity (eq. 3) we have
\begin{equation}
\label{veq}
v_{eq}=v_{lam} \biggl( \frac{L}{l_{eq}} \biggr) ^{D-2}
\end{equation} 
for the fractal velocity in the magnetic equator of the star,
while, as we have seen before,  eq. (3) gives $v_{pol}=u_{t}$ for the magnetic
poles, 
\begin{equation}
v_{pol} =v_{lam} \biggl( \frac{L}{l_{pol}} \biggr) ^{D-2}
\end{equation} 
with $l_{pol}$ given by eq. (4).

Taking the ratio between the velocities $v_{pol}$ and $v_{eq}$,
we obtain
\begin{equation}
\label{asym}
\frac{v_{pol}}{v_{eq}}=
\biggl( \frac{B^{2}/8\pi + \rho \, v_{lam}^{2}/2}{\rho \, v_{lam}^{2}/2} \biggr) ^{D-2} .
\end{equation}
The relevant data for do the calculations can be found in
Timmes \& Woosley (1992). We will use a fractal dimension $D=2.5$ as in Blinnikov, Sasorov \& Woosley (1995) and references therein.

Figure 1 displays the percentage of asymmetry in the
velocity field for two white dwarfs with different
compositions: (a) $X(^{12}C)=0.5\,\,X(^{16}O)=0.5\,$, $\Delta \rho /\rho=0.426$,
$v_{lam}=2.33 \times10^{5} \,{\rm cm\,s^{-1}}$ and 
(b) $X(^{12}C)=0.2\,\,X(^{16}O)=0.8\,$, $\Delta \rho /\rho=0.415$, 
 $v_{lam}=0.415\times 10^{5} \, {\rm cm\,s^{-1}}$,  and fuel densities 
$\rho\sim10^{8}\,{\rm g\,cm^{-3}}$ which are encountered
by the front at a middle radius distance $\sim 8 \times 10^{7}\,{\rm cm}$
from the center of the star. We see that the asymmetry
is  very sensitive to the composition of the progenitor and is quite insignificant
for type (a) progenitor.  
If at these densities, magnetic fields as high
as $\gtrsim 6\times10^{8}$ G are present at this radius, the velocity field
has an asymmetry $\gtrsim 10\,\%$ for progenitors of type (b). 
With surface magnetic fields of the order of $10^{9}$ G, and  fields 
$\gtrsim 6\times10^{9}$ G at the interior, these progenitors  could suffer huge asymmetries $\gtrsim 120\,\%$ (see Fig. 1).
Progenitors with compositions $^{16}O\,^{20}Ne$ show even higher asymmetries
(see Ghezzi, de Gouveia Dal Pino \& Horvath 2001) because the laminar velocity in $^{16}O\,^{20}Ne$ is smaller than in $^{12}C\,^{16}O$.
Also, we note that for regions very close to the center of the star (at higher densities) the asymmetry percentages are smaller than $5\,\%$. 

When the flame reaches the lates stages of the flamelet regime,
at a density of $\sim 5\times10^{7}\,{\rm g\,cm^{-3}}$, the magnetic field
intensity drops to $\sim 10^{8}-10^{9}$ G, however 
our calculations indicate that the percentage of asymmetry is maintained or increased
(Ghezzi, de Gouveia Dal Pino \& Horvath 2000),
leading to an asymmetry higher than $\geq 10\,\%$ for a progenitor with 
$X(^{12}C)=0.2\,\,X(^{16}O)=0.8\,$ with fields higher than $10^{8}$ G at the 
surface of the star.

\section{Conclusions and Discussion}

We have found that an asymmetric field velocity caused by the presence of a dipolar
magnetic field during the fractal growth of the deflagration front of a type
 Ia supernova  can lead to the formation of a prolate remnant. 
The magnetic field introduces an effective surface tension 
in the equator of the white dwarf progenitor that reduces the velocity
of the combustion front, $v_{eq}$, with respect to the velocity at the poles,
$v_{pol}$, so that $v_{pol} > v_{eq}$ by a percentage of $10\,\%$ to $20\,\%$ 
for progenitors with a composition 
$X(^{12}C)=0.2\,\,X(^{16}O)=0.8\,$, $\Delta \rho /\rho=0.415$, and 
surface fields $\sim 10^{8}$ G (type (b), Fig. 1). 
This leads to prolate explosions along the magnetic poles.
  Lower magnetic field strengths have no detectable effects on the explosion.
 As only a small fraction of the observed white dwarfs must 
have magnetic fields as high as $10^{8}$ G, asymmetric explosions 
are not expected to occur very frequently.
If there is no transition to detonation the asymmetry will be probably maintained during 
the free expansion phase of the supernova remnant, because the expansion velocity is constant at 
this phase. 
Detonation in an SN Ia is still controversial, 
but if a transition to detonation occurs the remnant can be symmetrized 
very fast (Ghezzi, de Gouveia Dal Pino \& Horvath 2001). 

Recent spectropolarimetric observations have revealed a linear polarization
component in the radiation of very young SN Ia remnants,  
which suggests that prolate atmospheres with asymmetries $\sim 20\,\%$ are 
producing it (see Leonard, Filippenko, \& Matheson 1999, 
Wang , Wheeler \& H\"oflich 1997).
The model presented here offers  a plausible explanation for such observations
and provides new motivation for theoretical studies of supernovae involving magnetic fields.
However, its predictions must still be confirmed through fully numerical
simulations of the explosion in the presence of magnetic fields.

\acknowledgements
This paper has benefited from many valuable comments by the referee F. X. Timmes, 
and discussions with M. Diaz. C.R.G.,   
E.M.G.D.P. and J.E.H. have been partially supported by grants of the Brazilian
Agencies FAPESP and CNPq.

\vspace{9.5cm}


\begin{figure}[h]
\epsfxsize=15cm
\epsfbox{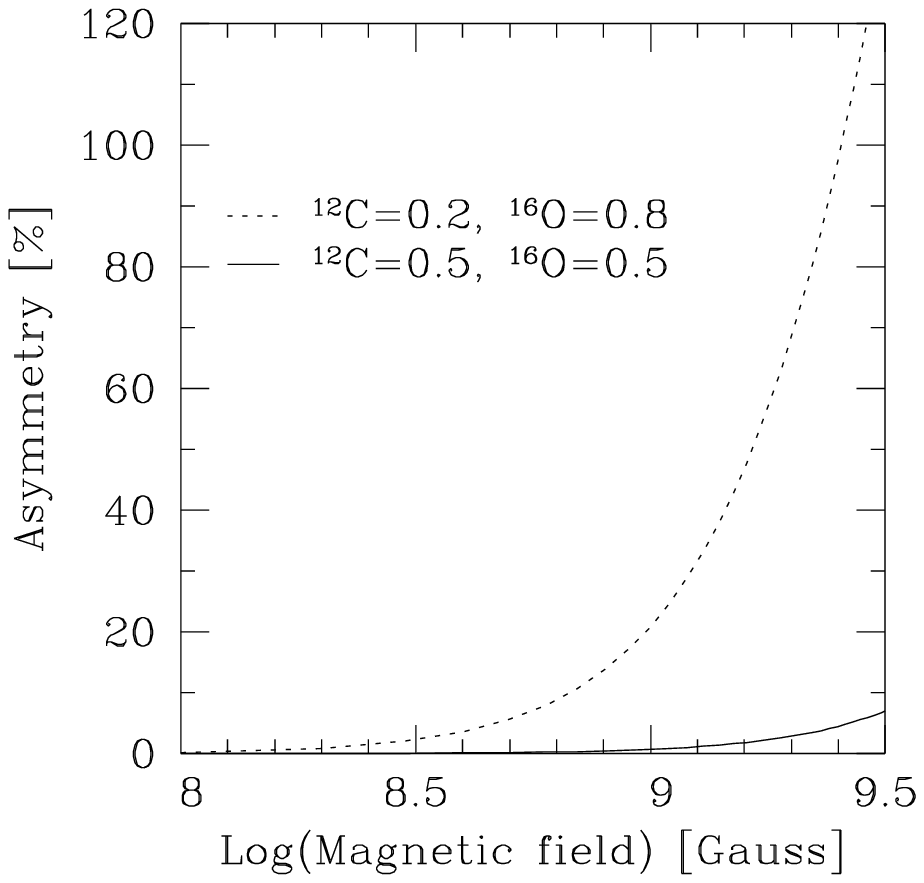}
\caption{Asymmetry percentage in the field velocity for two 
progenitors with initial compositions
$X(^{12}C)=0.5\,$ $X(^{16}O)=0.5\,$ and $X(^{12}C)=0.2\,$ $X(^{16}O)=0.8\,$, 
at $\rho \simeq 10^{8}$ ${\rm g\,\,cm^{-3}}$, as a function of the magnetic field strength at a radius $\sim 8 \,\times\,10^7$ cm. The asymmetry percentage is given by: 
$100 \times \biggl( v_{pol}/v_{eq}-1 \biggr)$ where $v_{pol}/v_{eq}$ is 
obtained by eq. (9).}
\end{figure}

\end{document}